# Beyond Flashcards: Designing an Intelligent Assistant for USMLE Mastery and Virtual Tutoring in Medical Education (A Study on Harnessing Chatbot Technology for Personalized Step 1 Prep)

Author: Ritwik Raj Saxena

Graduate Teaching Assistant; Research Advisee, Department of Computer Science, University of Minnesota, Duluth Campus.

Corresponding author: Ritwik Raj Saxena
Email: saxen130@d.umn.edu

**Abstract**

Objective: Traditional educational approaches, especially in the grueling medical basic sciences education paradigm, almost ubiquitously follow a one-size-fits-all model, neglecting the diverse learning styles and paces of individual students. I propose an intelligent AI companion which will fill this gap by providing personalized and adaptive learning experiences and which will provide on-the-fly solutions to students' questions in the context of not only USMLE Step 1 preparation but also the preparation of other similar examinations for local and international students in other countries, including PLAB Part 1 in United Kingdom, and NEET (PG) and FMGE in India.

Methods: I have harnessed Generative Artificial Intelligence (Generative AI) for dynamic, accurate, human-like responses and for knowledge retention and application. Users were encouraged to employ prompt engineering, in particular, in-context learning, for response optimization and enhancing the model's precision in understanding the intent of the user through the way the query is framed. The implementation of Retrieval Augmented Generation has enhanced the chatbot's ability to combine pre-existing medical knowledge with generative capabilities for efficient and contextually relevant support. Mistral was employed using Python to perform the needed functions.

Result: The digital conversational agent was implemented and achieved a score of 0.59 on a reference-based metric similar to Bilingual Evaluation Understudy (BLEU) and Recall-Oriented Understudy for Gisting Evaluation (ROUGE) scores.

Conclusion: My cutting-edge approach addresses a critical gap that exists in traditional medical basic sciences education by introducing an intelligent AI companion which specializes in helping medical aspirants with planning and information retention for USMLE Step 1 and other similar exams. Considering the stress that medical aspirants face in studying for the exam and in obtaining spontaneous answers to medical basic sciences queries, especially whose answers are challenging to obtain by searching online, and obviating a student's need to sift through bulky medical texts or lengthy indices or appendices, I have been able to create a quality assistant capable of producing customized, ad-libitum responses best suited to the user's needs.

**Keywords:** Intelligent conversation agent, chatbot, Generative AI, embeddings, in-context learning, prompt engineering, tokenization, Retrieval Augmented Generation, response optimization, conversational AI, Generative Pre-Trained Transformer, embedding-ada-002, Embedchain, medical boards examination, USMLE, PLAB Part 1, FMGE, NEET (PG), undergraduate medical education, medical students, education, medical education, medicine, international medical graduates.

## 1. Introduction

Generative AI is a branch of artificial intelligence focused on creating new data instances, such as images, text, or even entire pieces of content, based on patterns learned from existing data. It leverages techniques such as generative adversarial networks (GANs), variational autoencoders (VAEs), and autoregressive models to generate realistic and novel outputs.

GAN is a generative AI paradigm which generates lifelike images based on a user prompt. It utilizes adversarial modeling. It consists of a coupling of two models, a generator and a discriminator. The generator aims to produce lifelike samples, while the discriminator, which takes the generator's output as its input, attempts to distinguish between real and generated samples. When a discriminator fails to classify a sample produced by the generator as a generated sample, the most realistic possible output is assumed to be produced. For efficient training in a Generative Adversarial Network (GAN), the generator is trained first, and while it is trained, the discriminator is frozen so that the weights of the discriminator are not updated. Then both the models are trained simultaneously, as the cross-entropy loss is iteratively reduced, and the model weights are updated until the discriminator achieves the highest accuracy in distinguishing generated images from real ones and the generator is able to create the most convincing output possible, which is in the form of an image.

Large Language Models (LLMs) are a groundbreaking improvement in the arena of generative AI in particular and technology in general. LLMs inspire meaningful innovation in the development of useful software. LLMs like GPT (Generative Pre-Trained Transformer), BERT (Bidirectional Encoder Representations from Transformers) and LLaMA (Large Language Model Meta AI) are trained on a massive assortment of data. Being trained using enormous corpora and having multiple layers with an immensity of parameters and nodes enables LLMs to possess a wide set of capabilities, including an immense breadth of knowledge. Even so, they can suffer from inadequacies in expert knowledge areas. Their absolute capabilities stay unexploited when they are employed without being coupled with a tool that supplements their powers. They need to be enhanced so that they do not struggle when generating responses to complicated queries which require specialized understanding.

LLMs are endowed with the capability to predict subsequent tokens in a sequence, based on previously generated tokens and an input prompt. They operate on the tenets of neural networks, specifically leveraging multi-head attention mechanisms to grasp the contextual nuances embedded within input sequences. By employing attention, LLMs discern the saliency of individual tokens within the input. This lets them focus on the most pivotal expressions while processing information.

The iterative nature of LLM predictions underscores their incremental approach, as they proceed token by token towards generating a complete output. This incremental generation process imbues LLMs with a unique trait: the ability to offer solutions across a myriad of problem domains, owing to their zero-shot capabilities. Bestowed with extensive pre-training on colossal datasets, LLMs possess a broad yet nuanced understanding of language patterns and structures, rendering them formidable "jacks-of-all-trades" that can address diverse tasks.

However, to tailor LLMs to specific tasks and elevate them to the echelon of "subject experts," techniques like Retrieval Augmented Generation (RAG) and fine-tuning attain imperativeness. The architectural blueprint underpinning LLMs mirrors the avantgarde Transformers architecture. While various LLM variants exist, denoted by descriptors such as small, large, XL, and so forth, these designations correlate with the magnitude of parameters (parameter count) present within the model. The number and other properties of parameters are symbolic of the model's complexity and capacity and dictate its NLP prowess and computational demands.

Loading LLMs into memory presents a formidable challenge, owing to their prodigious size. Moreover, the generative nature of LLMs introduces an element of randomness into their outputs, ensuring that no two outputs for the same input are identical. However, while the transformative potential of LLMs is undeniable, a veil of secrecy often shrouds precise details regarding the training data and architectural intricacies, specifically of closed-source LLMs, leaving some aspects of their design enigmatic and inscrutable.

LLMs can spit out responses which may cause a user to think they are hallucinating – responses that contain unintended, inaccurate, or subjective details. Such skewedness is a consequence of LLMs being trained over vast, unprocessed web data. To redress this bias, LLMs can be augmented with tools like RAG, in-context learning, finetuning, transfer learning and prompt engineering. Equipped with such appurtenances, LLMs can fathom and make use of information with higher contextuality and respond with better targetedness.

In the realm of NLP, processing sequential input and generating sequential output become paramount for many of the most valued applications of NLP such as dialogue agents, machine translation, text summarization, image captioning, speech recognition, and time series prediction. The models which take sequences as inputs and out sequences in return are called sequence to sequence models.

Sequence to sequence models can broadly be categorized into two kinds, based on the way they are trained. Beyond the surface-level distinctions like parameter count, activation functions, architectural intricacies, contextual dimensions, scope of pretraining datasets, and programming language diversity employed during training of these models lie the often-underestimated subtleties of the training process itself. Such deep learning models can be classified into two overarching paradigms: the auto-regressive paradigm akin to GPT, and the auto-encoding paradigm reminiscent of BART (Bidirectional and Auto-Regressive Transformers).

Sequential data processing lies at the core of numerous applications within the domain of deep learning, encompassing realms such as natural language understanding, speech recognition, and time series forecasting. At the forefront of handling sequential data are three significant architectures: Recurrent Neural Networks (RNNs), Long Short-Term Memory (LSTM) networks, and Transformers.

### 1. Machine Language Translation

Modelos transformadores como o GPT-3 usam mecanismos de atenção. 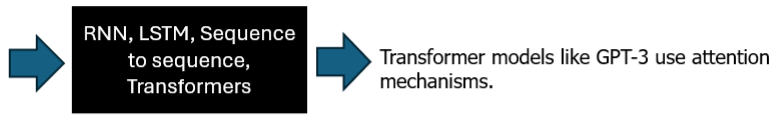 Transformer models like GPT-3 use attention mechanisms.

### 2. Text Summarization

I wish you all a wonderful and optimistic morning. 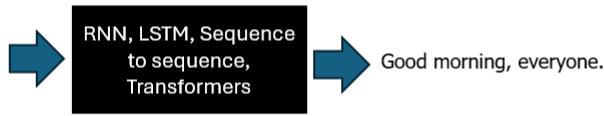 Good morning, everyone.

### 3. Chatbot implementation

What should a patient presenting with acute onset chest pain worsened by rapid left arm movements in the context of prior myocardial infarction and coronary artery bypass grafting be raising clinical suspicion of? 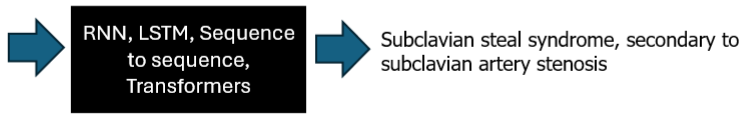 Subclavian steal syndrome, secondary to subclavian artery stenosis

Figure 1: The figure illustrates some applications of neural networks in natural language processing

*1.1. Autoregressive Models*

"Time discovers truth." – Seneca. If an artificial intelligence model can infer serial inputs, it can provide solutions to many sequential problems. Autoregression is a statistical methodology employed in time-series analysis. It is based on the assumption that the current value of a variable is a function of its past values. Autoregressive models employ methods that are akin to techniques visible in nature. While the specific implementations may vary, the underlying principles used by different autoregressive models share common characteristics.

Autoregressive models are designed to capture the probabilistic correlation between elements in a sequence and leverage this knowledge to predict the next element in the sequence. An autoregressive model predicts the value of a time series (or token in the case of language modeling) at a given time point based on its own previous values. Mathematically, an autoregressive model of order *p,* or AR(*p*), predicts the value of $y_t$ (the variable at time *t*) as a linear combination of its *p* previous values, plus an error term:

$$y_t = c + \sum_{i=1}^{p} \phi_i \cdot y_{t-i} + \epsilon_t$$

where $c$ is a constant (which is usually equal to 1), $\phi_i$ are the parameters of the model (are the coefficients that need to be learned from the data), $y_{t-i}$ are the previous values of the variable, $p$ is the number of lags, also called the lag window ($p$ is the number of past data points that are used to generate $y_t$) and $\epsilon_t$ is the error term at time $t$ (Kotu and Deshpande, 2019).

As shown, autoregressive language modeling aims to predict the probability distribution of the next token in a sequence based on the tokens that precede it. An autoregressive language model is trained to $P(x_t|x_{t-1}, x_{t-2}, \ldots, x_1)$, where $x_t$ is the token at time $t$ and $x_{t-1}, x_{t-2}, \ldots, x_1$ are the tokens that precede it in the sequence. During training, the model is fed with a sequence of tokens up to time *t-1*, and it predicts the probability distribution over the vocabulary for the next token $x_t$. Such models are typically trained using maximum likelihood estimation, where the goal is to maximize the likelihood of observing the true sequence of tokens in the backdrop of the model parameters.

For instance, an autoregressive model which is trained on several English language statements and recognizes that the word "*am*" has a considerable probability of following the word "*I.*" Using this knowledge, it generates a new sequence that has "*I am*".

*1.2. Auto-encoding Models*

The concept of feature learning, also referred to as representation learning, encompasses a spectrum of methodologies that empower a machine learning model to determine the requisite representations crucial for feature detection and classification directly from unprocessed data (Lin *et al.*, 2023; Liu *et al.* 2020). It entails the utilization of algorithms to extract actionable patterns from raw data to generate representations that are more comprehensible and conducive to computational processing. In the case of large language models, these representations imply embeddings and vectors, and they provide information regarding the features of the text to the model. The creation of these representations is termed encoding, and the dimensions of the embeddings are usually different from that of the input data. Using these representations, decoders and downstream models are able to easily interpret the input data.

They can identify latent features using these representations. The underlying patterns observed from these embeddings inspire the architectures to be modeled on themselves, aligning their parameters in line with the embeddings. This is called finetuning. In tasks such as image classification and retrieval, representation learning assumes a central role. Deep learning models are able to efficiently encode information that is subsequently mapped onto a distinct subspace of a new dimension. Following this encoding, these representations become input to decoders, also called classifiers. The decoders harness the encoded information to train and to engage in transfer learning.

Auto-encoding involves representation learning as one of its chief features. Autoencoders are neural network architectures used for unsupervised representation learning. They compress input data, for instance, images and text, into a lower-dimensional latent representation (encoding) and then reconstruct the original input from this encoding. BART is a notable example of an autoencoder which performs this task (Lewis, M., *et al.*, 2019).

Autoencoders help encode high-dimensional data. A text autoencoder trained on a large corpus of news articles, for example, can help us generate headlines for news snippets. preprocess and tokenize a collection of news articles. To successfully accomplish this, the encoder compresses each article into a lower-dimensional latent vector. The decoder reconstructs each original article from this latent vector. Then a random latent vector is sampled from the learned distribution, the loss is minimized, and the decoder generates a sequence of words based on the latent vector with the maximum likelihood.

The encoder function in an autoencoder can be depicted as follows:

$$z = f(x; \theta_{ec})$$

Here, $z$ represents the latent variables obtained from the input data $x$, and $\theta_{ec}$ denotes the parameters of the encoder function, while f denotes the function which helps us find the latent variables, each corresponding to an input.

The decoder function can be represented as:

$$x_{reconstructed} = g(z; \theta_{dc})$$

Here, $x_{reconstructed}$ is the reconstructed input generated from the latent variables $z$, and $\theta_{dc}$ denotes the parameters of the decoder function, while g denotes the function which helps us find the reconstructed input using latent variables and the decoder parameters.

To find the loss (the objective function is supposed to minimize the loss), we use reconstruction loss, which measures the discrepancy between the original input $x$ and the reconstructed input $x_{reconstructed}$. The training objective is to minimize the reconstruction loss by adjusting the parameters $\theta_{ec}$ and $\theta_{dc}$ of the encoder and decoder functions, respectively.

In his 2016 paper, Doersch discusses variational autoencoders. According to Doersch, in variational autoencoders (VAE), the latent variable $z$ represents points in a high-dimensional space. $z$ can be sampled from a probability density function (PDF) $P(z)$ defined over the high-dimensional space (we shall call this $Z$). The encoder functions map points in the latent space to points in the input space which contains $x$. The latent variable $z$ is randomly sampled from $P(z)$ and $\theta$ (set of parameters or weights) is fixed. $f(z; \theta)$ becomes a random variable in the input space which contains $x$. For each sample $z$, the corresponding output $f(z; \theta)$ can vary. The objective is to optimize the parameter vector $\theta$ such that for each data point $x$ in the training set, the probability of generating $x$ under the entire generative process is maximized. Doersch represents this as maximizing the probability $P(X)$, (where $X$ represents the input space) which can be expressed as the integral over all possible latent variables $z$ of the conditional probability of $P(X)$ given $z$ multiplied by the probability density function $P(z)$:

$$P(X) = \int_{z \in Z} P(X \mid \mathbf{z}; \boldsymbol{\theta}) \cdot P(\mathbf{z}) \, dz$$

In lieu of $f(\mathbf{z}; \boldsymbol{\theta})$, we use $P(X \mid \mathbf{z}; \boldsymbol{\theta})$ to make the function probabilistic in nature. It is based on maximum likelihood estimation (Doersch, 2016).

VAEs excel at mapping each input to a probability distribution in a latent space. This latent space has a lower dimension compared to the input space. As against traditional autoencoders, which simply map the input to a single fixed point in the latent space, VAEs learn a mapping from input data to a probability distribution. The decoder in a VAE reconstructs the original data from a point in the latent space. The probabilistic nature of the latent space allows for more flexibility in this reconstruction. Sampling points from the learned latent space distribution helps VAEs generate entirely new data points that resemble the training data.

In a VAE, the encoder, after accepting the input data, lets it traverse a dense layer to generate latent variables applying the ReLU activation function, and produces the mean and log variance of the latent space distribution. The decoder network is a feed-forward neural network that takes the latent space representation as input, lets it cross a dense layer to generate latent variables applying the ReLU activation function, and creates outputs by employing a new dense layer which has a sigmoid activation function. The sampling function takes the mean and log variance of a latent space as inputs and generates a random sample by adding noise scaled by the exponential of half the log variance to the mean.

*1.3. A Unified View of Sequential Modeling: RNNs, LSTMs, and Transformers*

In sequential modeling data unfolds over time. Sequence to sequence models are vital for various tasks like language processing, speech recognition, and time series forecasting. This domain is dominated by two main paradigms, RNNs and Transformers. LSTMs are considered a distinct type of RNNs.

"Time will explain it all," quoth Euripides, sometime in the fifth century B.C. The incorporation of temporality in deep learning architectures helps in solving tasks that require learning and interpretation of sequential patterns. RNNs are the foundation of sequential models. They possess an internal memory in the form of hidden states. The current states of RNNs incorporate information from previous inputs through hidden states which empower them to utilize previous information for processing current information. This makes them adept at capturing short-term dependencies within sequences. RNNs can effectively capture temporal dependencies and patterns within sequential data, making them well-suited for tasks such as music control, language modeling, and sequence generation. They have also been extensively used for sentiment analysis (Saxena, 2024).

As new input data is processed, the activations of an RNN's hidden units capture and encode information about the current input context, which can influence the network's behavior and predictions over subsequent time steps. RNNs consist of feedback connections which enable them to store representations of recent input events as activations. The feedback connections in RNNs enable the propagation of information from one time step to the next, creating a dynamic memory of recent inputs. According to Hochreiter and Schmidhuber, this can be referred to as short-term memory (Hochreiter & Schmidhuber, 1997). Unlike the long-term memory embodied by the slowly changing weights of the network, this short-term memory mechanism allows RNNs to retain information about recent inputs and incorporate them into subsequent computations.

However, a significant challenge arises with RNNs in the form of the vanishing gradient problem. As sequences lengthen, the influence of earlier inputs diminishes exponentially, hindering the network's ability to learn long-term dependencies. To combat the vanishing gradient problem gating mechanisms emerged. LSTMs introduce a concept called a cell state, which acts as a gated memory. This memory cell or LSTM cell is a layer of neurons, with each neuron having a hidden layer and a current state. Information can selectively flow through these gates, allowing LSTMs to remember relevant information over longer sequences and overcome the vanishing gradient problem.

According to Jurafsky and Martin, LSTMs can eliminate unneeded information from the context while adding information which will likely be required to take decisions later. LSTMs manage context using a context layer to the model. The neural units of LSTMs utilize gates to control the flow of information into and out of the units that comprise the network layers. These gates are implemented by employing fresh weights which are applied sequentially to the current input layer, the preceding hidden layers, and the preceding context layers. The first gate, that is the forget gate, selects whether the information coming from the previous timestamp is to be remembered or is irrelevant and can be forgotten. Using the input gate, which is the second gate, the memory cell endeavors to commit fresh information to itself from the data that is fed as input to this cell. Using the third gate, the cell passes the revised information from the present timestamp to the subsequent one. This constitutes a single cycle or a time step of LSTM.

LSTMs, with their intricate architecture comprising memory cells and gating mechanisms, are well-suited for tasks that demand the network to retain and utilize contextual information over extended sequences, such as language translation and speech recognition. However, training LSTMs can be challenging due to their complex structure.

Similar to LSTMs, GRUs (Gated Recurrent Units) employ gates to regulate information flow. They have a simpler architecture with fewer parameters, making them computationally more efficient. GRUs feature a single gate called update gate which regulates the flow of information into the memory cell. This streamlined architecture makes GRUs easier to train and faster to execute compared to LSTMs. While GRUs may not be as effective as LSTMs in storing and accessing long-term dependencies, they demonstrate proficiency in tasks requiring rapid adaptation to new inputs. Examples include time series forecasting and applications where quick learning is crucial. In the decision between LSTM and GRU architectures, there is no one-size-fits-all solution. The choice depends on the specific requirements of the task at hand. For complex tasks demanding the retention of long-term dependencies, LSTMs may offer superior performance, while for simpler tasks which prioritize fast adaptation and ease of training, GRUs are more effective.

RNNs, LSTMs, GRUs, and their other variants like BiRNNs (Bidirectional RNNs), BiLSTMs (Bidirectional LSTMs) and BiGRU (Bidirectional GRUs) can be combined with other models like hidden Markov models (HMMs) based on the task that needs to be accomplished. A bidirectional model couples two models which employ similar architectures. One of these models takes the input in a forward direction while the other carries it in a backwards direction. Bidirectional models process sequential information in both directions (forward and backward) simultaneously. This allows them to capture contextual information from both the past and future, leading to improved performance on tasks like sentiment analysis and machine translation. Bidirectionality enhances the quantity of information available to the network. This helps in better training by increasing the context available to the algorithm. A bidirectional model, for instance, can have better understanding of which words must immediately follow and go before a particular word in a statement. Another approach to capture complex sequential relationships is stacking multiple RNN layers. Each layer receives the output from the previous layer, allowing the network to learn hierarchical representations of the data. This can be particularly beneficial for tasks with intricate temporal dynamics. Like graph neural networks (Saxena and Saxena, 2024), stacked RNNs learn hierarchical representations.

Transformers are models which enable users to significantly reduce the requirements of computational resources, environmental impact, and the resources required for training models from scratch. Covering a diverse range of tasks across various modalities, transformers can support natural language processing tasks like text classification, question answering, and text generation, as well as computer vision tasks such as image classification, object detection, and segmentation. They excel in audio-related tasks like automatic speech recognition and audio classification, while also offering solutions for multimodal applications like table question answering, video classification, and visual question answering.

"Attention is vitality." — Susan Sontag. "The true art of memory is the art of attention." — Samuel Johnson. "The simple act of paying attention can take you a long way." — Keanu Reeves. All these sayings about attention validate its utility in NLP. Attention is the vital thread that runs through the preponderance of the most efficient modern LLMs. Attention dramatically improves the contextual memory capacity of an LLM and currently stands unparalleled in its effectiveness. Transformers architectures, unlike RNNs, apply self-attention instead of relying on recurrence. While gating helps LSTMs forget insignificant pieces of information and retain significant ones, self-attention enables the model to assess the significance of various input tokens at the instance they make prediction.

In this manner, transformers can capture long-range dependencies more efficiently, even for very long sequences, without requiring sequential processing. A transformer is composed of feedforward neural networks. Transformers consist of two layers of encoders and two layers of decoders. As it processes data in parallel without recurrent connections, the transformer architecture excels at capturing dependencies between elements, regardless of their positions in the sequence, making it particularly potent for tasks involving long-range dependencies. Attention mechanisms allow transformers to attain a long-term memory, theoretically granting an infinite reference window if there are sufficient computing resources.

Both the encoder and a decoder in a transformer begin by sending the input to the word embedding sub-layer. (The output representation of the encoder stack in a transformer is a complex data structure containing information about the entire input sequence, which is sent as input to the decoder). After creating input embeddings, the same layer incorporates positional information in them by adding positional embeddings to input embeddings, which implies that positional embeddings have the same dimension as the input embeddings. Positional information is incorporated in the input embeddings to compensate for recurrence that RNNs introduce into the input data (Vaswani *et al.*, 2017).

The next comes the encoder sub-layer, which maps all input sequences into an abstract continuous representation that contains the learned information about the input sequence. This is done through the attention mechanism. Positionwise fully connected feed-forward networks are applied to the ouput of the encoder layer. They involve a linear, followed by a non-linear (ReLU), followed by another linear transformation. Residual connections and layer normalization are applied to both the encoder layer and the positionwise feedforward networks. In each layer of the decoder, an additional sub-layer, which performs multi-head attention on the output of the encoder, is used.

In the context of attention, three vectors, query, key, and value, are important. The three kinds of vectors are generated by applying linear transformations, using query projections, key projections, and value projections (which are three unique weight matrices), respectively, to the input data. Since the query, the key and the value have the same base, that is, the input data, the attention mechanism is called self-attention – the input pays attention to itself.

The query vector represents the information that a model seeks to retrieve in the input sequence. It acts as a reference for comparing against the keys to determine the relevance of different parts of the input sequence. The attention mechanism uses the query vector to calculate attention scores, indicating how much focus should be placed on each element of the input sequence. The key vector represents the information against which the query vector is compared to determine relevance. It provides a set of reference points that the model uses to evaluate the similarity between the query and different parts of the input sequence.

The key vector helps compute attention scores by measuring the similarity between the query and each element of the input sequence. The value vector represents the actual information associated with each element of the input sequence. It stores the content of each input element, which the model can retrieve based on the attention scores obtained from the query-key comparisons. To illustrate, if we equate attention to an information retrieval search, the query represents the search terms entered, the key embodies the representation of each of the disparate pieces of information or text and is compared against the query to obtain the relevance of a particular piece of information, and the value vector includes the details of all pieces of information. The query vector represents the search query, the key vectors represent the content of the pieces of text in the database searched, and the value vectors contain the actual details associated with each piece of information.

In case of a simple input statement "He eats vegetables.", the linear transformation with the query vector will reveal "he" as the query, which can also be termed as the subject. The linear transformation with the key matrix reveals "eats", which is the most related word to "he" in the statement. And the value for "he eats" shall be "vegetables" (as the answer to the question "What does he eat?" or what is the value of the variable "he eats".)

A transformer calculates the output vector by multiplying the attention weight with the vector of values. Hence, the output of a transformer becomes a weighted sum of the values. The query projections and the key projections exist in such a way that the query vector and key vector have the same dimensions. The attention weight is calculated by finding the scalar dot product of the query vector and the matrix transpose of the key vector, dividing it by the square root of the dimension of the key or the query vector to scale it, and then performing a softmax operation on the resulting value. The output vector, therefore, has the same dimensions as the value vector. This procedure is referred to as scaled dot product attention.

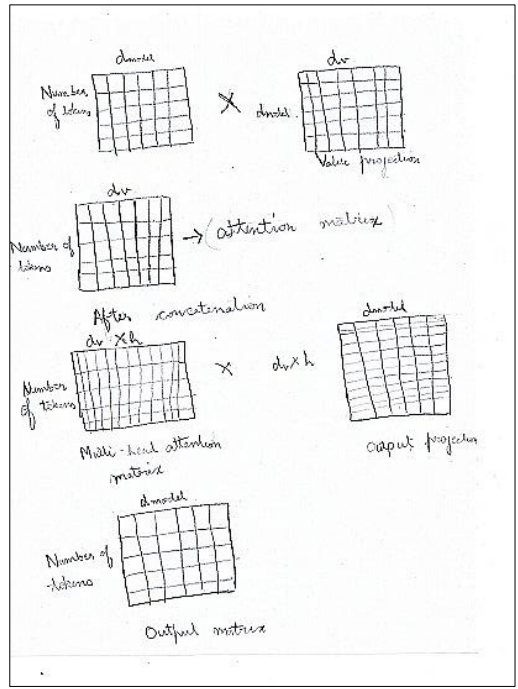

Figure 2: The dimensions of the output matrix of a transformer are the same as those of the input matrix. Here $d_{model}$ implies the dimensionality of the model, or, in other words, the size of the embedding vector for the model; $d_v$ implies the dimensionality of the value vector; h implies the number of attention heads.

The transformer applies multiple linear projections of queries, keys, and values on each input. Each linear projection is called an attention head. As scaled dot product attention is applied to these projected versions in parallel, output vectors are also generated in parallel. They are concatenated and projected once again with an output projection to obtain the final output. Here, the number of rows of the output projection is equal to the number of heads multiplied by the dimension of the value vector, while its number of columns corresponds to the size of the transformer's embedding, or the dimensionality of the architecture. The operation results in an output vector which is of the same dimensions as the input vector (Figure 2). This phenomenon of calculating different scaled dot product attention output vectors and then concatenating them to get the final output is termed as multi-head attention.

Unlike encoders, decoders use masked self-attention in addition to multi-head attention. Masked self-attention keeps the model from focusing on tokens that are positionally placed ahead of the token whose attention scores are being calculated. It is also present in decoder-only transformer models. This makes the operation autoregressive in nature since the only value that a token's attention score is dependent on is the attention score of the past tokens.

Training complex models like transformers can be computationally expensive. Parallelization comes to the rescue, allowing us to distribute the workload across multiple processing units (GPUs, TPUs) for faster training. Data parallelism divides the training data into smaller batches and distributes them across processing units. Each unit trains the model independently on its assigned batch, and the gradients are averaged for weight updates. Model parallelism divides the model itself across processing units. Different parts of the model are assigned to different units, which communicate with each other to complete calculations. This is advantageous for very large models.

In this way, as compared to RNNs and LSTMs, transformers exhibit superior handling of sequence length and parallelization, enabling efficient training, and heightened scalability. The time complexity per layer in transformers is dependent on the square of the length of the sequence (the input is divided into sequences, each of which contains a fixed number of tokens) and on single power of the dimensionality of the model, unlike RNNs where the time complexity per layer is dependent on the square of the dimensionality of the model and on a single power of the length of the sequence. Transformers have larger model sizes. They accommodate complex tasks. They have shown much better performance in generative tasks and show even more effective performance with effective pre-training, finetuning, transfer learning, and in-context learning. RNNs offer higher interpretability because of their clear temporal course. Transformer architectures require more effort for interpretation, albeit attention visualization techniques provide insights into decisions that the models make.

*1.4. LLMs, and their Advantages and Limitations*

LLMs are built using architectures which can handle sequential data. These architectures include RNNs and transformers. LLMs have two main characteristics: first, they have sophisticated architectures with millions to billions of parameters, a large number of layers, a high dimensionality, and a massive number of neural units. Second, they are trained on copious quantities of data, making them capable of handling many problems.

The ability of LLMs to grasp nuances of human language and their vast training on text data make them attractive tools. LLMs excel at grasping the context within a query, something termed in-context learning. LLMs can pinpoint the most relevant keywords and produce relevant outputs by efficiently analyzing surrounding words and sentences. They can even handle ambiguity, interpreting user intent in confusing queries and extracting keywords that fit the specific scenario. Their capability to generalize allows them to adapt to slightly different phrasings and extract keywords effectively even for novel queries. This prowess can be further enhanced by fine-tuning them on specific domains or datasets, fostering continuous learning and improved accuracy over time.

However, LLMs also come with limitations. While powerful, they may not always align perfectly with a user's specific needs. Outputs generated by LLMs based on learned patterns from training data might be technically accurate but may sometimes miss the user's intended meaning, especially when the user employs domain-specific terminology in the query. LLMs are heavily influenced by the data they are trained on. Biases and limitations within that data can be reflected in their outputs, potentially leading to struggles with queries not well-represented in the training corpus. Their computational cost also needs consideration, as running LLMs can be expensive for real-time applications where expeditious training times, low latency and high throughput may be crucial. LLMs often lack transparency in their keyword selection and output generation process. They might produce pertinent outputs, but explaining why specific choices were made can be challenging due to limited interpretability.

*1.5. Long-context LLMs*

When sequence to sequence models can handle extended contexts, like entire documents, they may be termed as long context LLMs. Apart from interacting with long documents and being able to answer questions with regards to the contents of those documents, long context LLMs can help with generating aggregated summaries of multiple documents, managing long-horizon agent tasks (where an LLM can process a high volume of text inputs and can find useful information by processing a massive history), and having the ability to understand and handle repository-level code (Li *et al.*, 2024).

Li *et al.* have brough the reliability of LLMs in understanding lengthy contexts under scrutiny. They articulate that despite advancements in the fields of NLP and LLMs, accuracy rates of LLMs vary. Existing evaluation methods, such as perplexity and synthetic tasks, are not able to sufficiently capture real-world context understanding of LLMs. Li *et al.* assessed various LLMs using their own benchmark, LongICLBench, and concluded that the accuracy of thirteen state-of-the-art LLMs is poor when handling long contexts.

Their findings underscore the ongoing challenges in developing LLMs capable of comprehensively understanding and accurately processing lengthy contexts, urging further research and refinement in this domain. Apart from enhanced prompt-engineering, better in-context learning, and finetuning on specific datasets, traditional RAG as well as its improved variants offer a strong solution to such problems that users encounter when using LLMs, especially in long context settings.

*1.6. Variability in LLM Responses*

The variability in responses generated by LLMs to same or very similar inputs can be attributed to several factors. Firstly, techniques like finetuning an LLM change its weights and parameters, leading to different output to the same input. Secondly, LLMs utilize sampling techniques during text generation, resulting in different outputs even with the same input prompt. Thirdly, stochastic optimization algorithms such as stochastic gradient descent (SGD) introduce randomness during model training by updating parameters based on randomly selected data batches. Fourthly, for largely similar inputs, the complex architecture of LLMs, coupled with interactions among numerous parameters, magnifies small input differences, leading to diverse outputs. Fifthly, regularization methods like dropout and weight decay add randomness during training to prevent overfitting, influencing the model's behavior during inference. Internal states unique to LLMs can vary between runs, impacting decision-making and contributing to response variability. Lastly, tokenization and subword encoding methods affect how text is represented and processed. This further enhances variability in model outputs.

At times, the same LLM, without finetuning, context variation, or retraining on different datasets, can display different outputs to the same input, even after clearing all previous data. That is because LLMs typically include a process known as stochastic noise where some randomness is injected into the system solely to ensure variance in the possible output. Stochastic noise refers to random fluctuations deliberately and strategically introduced during the training process. This noise helps the neural network to break free from the local minima of the task it has been initially trained on and allows the network to adapt its weights and biases more readily to learn a new task. Without stochastic noise (or other randomization factors), if a neural network is in the same initial state and the input and weights remain unchanged, the network should consistently produce the same output for the same input. Temperature is another parameter which controls randomness. The higher the temperature, the less deterministic and repetitive and the more creative and unique an LLM's responses would be. An additional setting in LLMs which controls their randomness is Top P. Top-p is a sampling technique which chooses from the smallest possible set of words whose cumulative probability exceeds the probability $p$ as configured in the model's code.

*1.7. Benefits of Using Open-source LLMs*

Open-source LLMs are those LLMs which are accessible without paying a cost to the license-holders, the owners, or the creators of the LLMs. These LLMs can be finetuned by anyone and tailored for a large variety of tasks. Utilizing open-source LLMs has numerous advantages for individuals and businesses. One benefit lies in cost reduction, as there are no licensing fees involved, leading to lower initial and ongoing expenses. Companies can deploy these models freely, contributing to direct cost savings. Open-source LLMs offer increased control over proprietary data, which enhances data security and privacy. This independence from vendors also reduces dependency and associated costs. The transparency of code and the ability to customize language models empower organizations to tailor solutions to their specific needs by employing finetuning and transfer learning.

Open-source LLMs are also beneficial for their creators, for they allow public experimentation and research on the architectures, which enables new insights on how to improve their performance. Leveraging active community support encourages innovation and facilitates the incorporation of new features through community contributions, which are invaluable in dynamic environments. Finally, adopting open-source LLMs aligns with efforts to reduce the environmental impact of AI, contributing to sustainable practices.

While OpenAI's GPT series, exemplified by ChatGPT, initially grabbed attention with its impressive capabilities, a critical shift is underway. Proprietary models come with high costs and are often not compatible with finetuning and transfer learning because of their blackbox character and massive sizes and immense number of parameters which increase their complexity. Owing to this, these models lack the ability to leverage data in specific, custom formats like .py files. In response to this, a growing trend to embrace the smaller, open-source LLMs is emerging. Such models offer greater flexibility, are amenable to finetuning and transfer learning, and can be amalgamated with RAG pipelines, letting users to allow these models to operate on their own data and produce actionable results.

*1.8. ChatOps and Compound AI Systems*

In the realm of modern software development and IT operations, a holistic paradigm emerges at the convergence of DevOps, MLOps, LLMOps, and ChatOps. This integrated approach transcends traditional boundaries, encompassing software development, agile methodologies, process optimization, and automation. DevOps stands at the intersection of software development, software process enhancement, agile framework and team-based software development, and automation in IT operations. Coding is merely a tiny part of it, and its essence lies far beyond code creation. Serving as a methodology, DevOps orchestrates a fusion of practices and tools to streamline and accelerate the software lifecycle. It leverages practices and tools to bolster an organization's capacity to deploy applications and services swiftly, outpacing traditional software infrastructure management paradigms. DevOps helps in accelerating delivery and in improving adaptiveness towards software engineering bottlenecks in the current digital landscape.

Embedded within the DevOps framework are MLOps and LLMOps. MLOps leverages machine learning models as foundational elements for AI-oriented applications, helping developers build their machine learning code into deployable artifacts. MLOps streamlines the operational aspects of AI development, from data management to model deployment and monitoring. Within that, LLMOps focuses on the organization and optimization of large language models, ensuring their seamless integration into the application development pipeline. Using LLMs to develop superior, expedited, and high-performing software applications for an above-board user experience constitutes LLMOps. The idea of ChatOps revolutionizes communication and collaboration through realtime chat platforms.

By integrating LLM-based conversation tools, chat clients, and chatbots, which synergistically integrate existing techniques in deployable applications in order to automate workflows, ChatOps melds innovation with efficiency and precision, enhances user experience, and facilitates streamlined incident response and task management. The concept of ChatOps brings the creation of novel conversational AI agents into focus. The notion of CI/CE/CD (Continuous Integration/Continuous Evaluation/Continuous Deployment) has gained prominence almost parallelly alongside the concepts of DevOps, MLOps, LLMOps and ChatOps. CI/CE/CD is important to maintain and to manage the lifecycle of applications which are built with LLMs as their underlying base.

The evolving landscape of AI marks a departure from conventional methodologies and is shifting towards exploring the development of Compound AI Systems. Being removed from the monolithic frameworks of the past, these systems encapsulate modularity and collaboration. They integrate diverse LLM applications with open-source tools to construct intricate pipelines. Unlike their closed counterparts (traditional AI models and systems which work in a proprietary manner), these systems integrate open-source components, thereby prioritizing flexibility and facilitating innovation in AI. While the foundational LLM remains consistent across various iterations, the differentiation lies in the crafting of prompts and the assimilation of contextual information. A compound AI system can be delineated into distinct components with specialized functionalities, all of which are parts of the deployed artifact.

An RAG system forms one of the components of a compound AI system. It employs the injection of new contextual data which is used to augment the answers to user queries. The injection of contextual data enables nuanced user query comprehension, the application of more human-like reasoning and thought generation to delve beyond superficial understanding typical of traditional AI systems. With a focus on allowing the setup to introspect, RAG systems enable Agent Assist setups, where AI agents are used to assist human consumer service agents. RAG systems can be incorporated with user-facing dialogue models (chatbots that users directly interact with) to trigger the development of robust conversational AI models as well as CoPilot architectures. Multi-Agent Problem Solvers harness the collaborative potential by inputting the intermediate outputs from specialized agents into each other to form a well-wrought, detailed, and comprehensive final output. Each agent possesses a unique toolset and has a specified function. This enables a smooth synthesis of solutions that transcend capabilities of individual models and boost those of humans.

Conversational AI systems epitomize the automation of human-computer interaction (HCI). They inspire natural chat within the digital fabric. These systems, powering virtual assistants and chatbots, excel in automating cyclical tasks while maintaining conversational memory and organic conversation generation. They leverage underlying RAG systems and Multi-Agent Problem Solvers to augment their capabilities and provide more complete, useful, and contextually relevant interactions. CoPilot Systems characterize the concept of human-in-the-loop (HITL) interfaces, collaborating with human agents to solve problems with predefined solutions. HITL models require human interaction and incorporate human aid to take correct decisions. CoPilots are equipped with all-encompassing tools, data, reasoning, and planning functionalities. They tailor their assistance to specific domains. They portray an advanced understanding of the human environment. This differentiates them from other systems as they are able to provide exceptionally valuable assistance.

*1.9. Chatbots: A Brief History and Various Types of User-facing AI Dialogue Agents*

Generative AI is a branch of artificial intelligence focused on creating new data instances, such as images, text, or even entire pieces of content, based on patterns learned from existing data. It leverages techniques such as GANs, VAEs, and autoregressive models to generate realistic and novel outputs. In the realm of Generative AI, conversational AI is a subset that specifically targets generating human-like conversations in natural language. Conversational AI systems aim to understand natural language input from users and generate appropriate responses, mimicking human conversation patterns and context. Chatbots are quintessential conversational AI agents that are commonly used today.

The history of chatbots can be traced back to the mid-1960s with the development of ELIZA by Joseph Weizenbaum at MIT, a rudimentary program designed to simulate a Rogerian psychotherapist. ELIZA utilized pattern matching and delivered open-ended responses. While it could not mimic true sentience, it nevertheless sparked a new era in Human-Computer Interaction (HCI). Following in ELIZA's footsteps, PARRY emerged in the 1970s as a chatbot designed to mimic a person with paranoid schizophrenia. Although it hallucinated, PARRY was a significant contribution to the field of AI. As a tool, it helped researchers study certain aspects of human psychology.

In 1983, a chatbot, Racter, was invented by Chamberlain and Etter. It deviated from the script-based approach of its predecessors. Racter's approach of pragmatics-driven language generation allowed it to produce original prose and engage in literary conversations. The late 1990s and early 2000s ushered in a new era of chatbot development with the rise of projects like ALICE and SmarterChild. ALICE, a more sophisticated text-generation agent, demonstrated improved coherence in responses while expressing human-like quirks like Racter did.

SmarterChild, a virtual assistant operating on instant messaging platforms, offers users information like weather details. Though not without its limitations, SmarterChild's accessibility endeared it to a wide audience. Similarly, Cleverbot, another chatbot marketed for its ability to learn and adapt through conversation, became a popular tool for casual interaction despite occasional nonsensical responses.

The 2000s saw unparallelled advancements in AI, machine learning, and cloud computing, and this led to a surge in interest and investment in chatbot technology. Companies began deploying chatbots for customer support, virtual assistants, and personalized recommendations. Platforms like Microsoft's Clippy, Apple's Siri, and IBM's Watson exemplified the growing capabilities of chatbots in understanding natural language and executing tasks. These advancements in chatbot technology laid the groundwork for the sophisticated virtual assistants and conversational AI like ChatGPT that we encounter today.

Modern chatbots are conversational AI agents which serve as virtual assistants, customer service representatives, or interactive companions in various applications. They interpret user queries or messages, process them using natural language understanding (NLU) techniques, and generate appropriate responses using natural language generation (NLG) methods. NLU focuses on deciphering the meaning and intent behind user messages by analyzing the linguistic structure and semantic context. It involves tasks such as entity recognition, intent classification, and sentiment analysis, aiming to extract actionable information from the user's input. NLU techniques employ various machine learning and natural language processing algorithms to parse and understand user queries, enabling the system to grasp the user's intentions accurately. It is a key component of machine translation.

NLG is responsible for crafting coherent and contextually relevant responses based on the insights gathered from NLU. NLG techniques leverage learned language patterns and semantic knowledge to generate human-like text that effectively communicates information and responds to user inquiries. NLG systems may employ rule-based approaches, template filling, or more advanced deep learning models to produce responses that are grammatically correct, contextually appropriate, and engaging to the user. By synthesizing information obtained from NLU with additional knowledge sources, NLG ensures that the generated responses meet the user's expectations and effectively address their queries or concerns.

Chatbots can be classified into various types, each with distinct characteristics and functionalities tailored to different use cases. One common classification divides chatbots into rule-based, retrieval-based, and generative types. Rule-based chatbots operate on predefined rules and decision trees, offering scripted responses determined by programmed logic. Retrieval-based chatbots retrieve pre-defined responses or templates from a database based on the similarity between user input and stored patterns. Menu or button-based chatbots, where users interact by selecting predefined options presented in menus or buttons. These chatbots offer a structured and guided user experience, suitable for scenarios where simplicity and clarity are prioritized. They are also called scripted chatbots.

AI-powered chatbots leverage advanced machine learning and natural language processing techniques to interpret user inputs and generate more nuanced and contextually appropriate responses. Generative chatbots are a subcategory of AI-powered chatbots. These chatbots utilize deep learning techniques like RNNs to dynamically generate responses based on the input context, allowing for even more flexible and contextually relevant interactions. They are also called contextual chatbots. Some of them are voice-enabled in nature.

Generative chatbots are of many types. Service action chatbots and social chatbots are two kinds of generative chatbots worth elucidating. Service action chatbots, also known as task-oriented or transactional chatbots, are focused on assisting users with completing specific tasks or actions. These chatbots are typically deployed within service-oriented environments, such as customer support, e-commerce platforms, or enterprise applications. Their primary goal is to streamline user interactions by providing quick and efficient access to information, services, or resources. Social chatbots are designed to engage users in more open-ended and conversational interactions, often with the goal of building rapport, entertainment, or emotional support. These chatbots simulate human-like conversation and aim to create a sense of connection or companionship with users.

## 2. Related Work

In the domain of chatbots for education, a number of articles and approaches are available. Machine learning for developing conversational agents has been a unique area of focus for about the past two decades. Farhan *et al.* in 2012, have presented a study that describes a web-based chatbot named Pandora. It is an e-learning platform that aims to address the issue of delayed responses to student inquiries. The chatbot predicts future events based on internet keywords. It utilizes AIML-stored questions and answers, with human intervention required when the chatbot cannot provide a suitable response.

Niranjan *et al.*, in 2012, proposed a chatbot designed to enhance the efficiency of question answering processes. They have described interdependent components within the architecture of the chatbot. At the forefront lies the interface engineered to emulate real-time interactions between students and the educational system. The tokenizer functions as a linguistic disassembler tasked with segmenting the student's inquiry into discernible tokens. The stop words eliminator winnows out extraneous linguistic elements (stop words) from the query corpus to reduce search complexity.

Central to the system's functionality is the Bayesian Module which uses the Naïve Bayesian algorithm. It narrows down potential answers until the most pertinent response is discerned. The paper also discusses three foundational databases. The category database is composed of various knowledge categories and the most common words associated with each category. The question database stores a repository of questions categorized under each knowledge domain. The answer database houses responses to queries cataloged within the question database. Then the Java speech engine-powered text-to-speech converter is encountered. A human agent referred to as the tutor assumes an administrative mantle, overseeing database management and curating question-answer repositories.

Colace *et al.*, have suggested a comprehensive AI-based conversation model comprising four principal modules: the front-end module, the back-office module, the knowledge base module, and the e-learning BOT Module. The front-end module serves as the user-facing interface, facilitating seamless interaction across a myriad of devices including tablets, smartphones, and PCs, thereby ensuring widespread accessibility and user engagement (Colace *et al.* 2018).

The back-office module operates in the background as it manages essential operations like business logic and data storage in concert with the knowledge base. The knowledge base module stores information and knowledge assets and processes user data and the learning object of a user. The linchpin of the system lies in the E-learning BOT Module, a multifaceted engine geared towards optimizing user interactions and facilitating dynamic dialogue. It monitors conversation logs, keeps track of dialogue dynamics, identifies ambiguous queries, implements context awareness, infers queries and mapping them to solutions using a Latent Dirichlet Allocator, constructs an ontology, handles user queries, and maps learning object metadata instances into the ontology.

Xu *et al.*, in 2017, developed a chatbot using conversation data scraped with the Twitter public API. They employed both the Streaming API and the Search API to capture tweets mentioning the selected brands. They removed non-English requests, requests containing images, and all Twitter mentions from both the training and testing datasets. Following data cleaning, the text data underwent tokenization, wherein a vocabulary comprising the most frequent 100,000 words in the conversations was constructed. Word embeddings were generated using the collected corpus to train word2vec models, wherein each word in the vocabulary was represented as a 640-dimensional vector. To develop the chatbot, LSTM networks were trained with stochastic gradient descent and gradient clipping for optimization.

Ranoliya *et al.*, in 2017, introduced a chatbot for university related frequently asked questions (FAQs). The chatbot developed in this research specifically caters to the educational domain, wherein users, including students and parents, can seek information pertaining to college admissions, institutional details, and other academic-related inquiries. A user query is directed to the AIML check module to ascertain whether the input conforms to AIML script standards. AIML responds to such queries through predefined templates. This process is delineated into three distinct stages: user submission of the query, processing of the query to align with predefined developer-specified formats, and subsequent pattern matching between the user's input and the stored knowledge patterns. Ultimately, a pattern-based response is generated and presented to the user to address their query effectively. Sharma *et al.*, earlier this year (2024), published a survey which summarizes various types of machine learning chatbots.

## 3. Background

While not many United States Medical Licensing Examination (USMLE) Step 1 aspirants face challenges accessing quality study materials, guidance, and expert support, international medical graduates aspiring to clear USMLE Step 1, particularly those preparing for the examination from remote and underserved areas, do not have it easy. International Medical Graduates (IMGs) not only have a poorer chance compared to US medical graduates (USMGs) of called for an interview and being admitted into a residency program, but also require much higher scores in USMLE Step 1, and therefore much more rigorous preparation, to enter a residency program in US (Murillo Zepeda *et al.*, 2022).

There is also a possible deficiency of cost-effective tailormade, one-stop, to-the-point, summarized study materials that capture the most important concepts of undergraduate medical education and provide solutions to basic sciences queries (basic sciences refers to the sum total of courses covered in medicine before a student sits for the USMLE Step 1, and includes anatomy, epidemiology, pathophysiology etc., which form the conceptual groundwork that a student must have a knowledge of before they embark on applying it in real world; it is a term specially coined to stand apart from the term clinical sciences, where applications of basic sciences like biochemistry, microbiology, pharmacology, etc. are studied in clinical settings – a curriculum which is implemented commonly after a medical student has competed in USMLE Step 1) without the need for looking up huge appendices or indices or sifting through a mountain load of pages, especially for IMGs.

In the realm of medicine and healthcare, there have been numerous AI-based conversation agents. For instance, there are chatbots that provide recommendations to people who are in remission from cancer. Emotional support chatbots, which monitor emotional health of patients and engage in conversations wherein psychological techniques are applied for support, are also available. Certain emotional support chatbots also offer breathing exercises. Some chatbots offer AI-based consultation services. Chatbots which send timely reminders to patients to take their medication are becoming a popular choice among patients of diseases such as dementia. Some chatbots help make optimum health choices and offer fitness recommendations, including charting out the best fitness routines after prompting users to input their physical conditions for healthy people as well as those undergoing physical rehabilitation regimens while or after suffering from a neuromuscular injury or disease. More sophisticated chatbots act as physicians' assistants that help in confirming specialists' diagnoses and providing independent diagnostic guesses based on patient data.

In the sphere of medical education and basic science eLearning, however, cases of AI-based conversational agents are few and far between. Ghorashi *et al.*, in their 2023 paper, have talked about potential chatbots in this domain. They have alluded to AI-powered interactive tutor and learning tools in basic medical sciences education. They have talked about potential chatbots that can act as interactive search engines in medical education, as tutors that can clarify challenging topics in medicine, and as virtual examiners that may offer realtime feedback, translating into valuable extra-classroom support for medical students.

Kaur *et al.*, in 2021, recommended using bespoke chatbots to help medical students in the study of medicine in Britain. They cited the rationale of limited teacher-student in-person contact during COVID-19 pandemic for introducing the conversational AI tool. They identified the use of chatbots as revision tools and as clinical stimulators/patient stimulators as two chief themes. They also identified chatbots as potential tools that may help standardize medical examinations. At the time of the publication of the paper, ChatGPT was new and not so well-known; so, they also suggested increasing awareness about chatbots among medical students as beneficial educative devices.

Xie *et al.*, (2023) examined the effect of AI chatbots on improving medical education and in serving as clinical assistants. They compared three online LLM-based chatbots in their performance in answering clinical scenarios. They demonstrated the possible practicality of such chatbots as agents that can help with self-directed learning and self-paced learning, as personalized education tools, and also as clinical decision-making assistants. Frangoudes *et al.* (2021) have explored the integration of chatbots in medical education, highlighting their applications, development process, evaluation metrics, and the need for further research, particularly within the European healthcare context. They conducted a review of papers to study chatbot acceptability, applicability of chatbots in the area of simulating virtual patients for medical students, as tools for patient education, and as course assistants for improving health professionals' curricula.

These, as well as many other studies, highlight that while there has been research, including surveys, reviews, and feasibility studies, into chatbots as medical education assistants as well as chatbots forged specifically as tools that augment the medical education paradigm, there have been no known conversational AI agents created specifically in the domain of medical basic sciences curriculum, explicitly in the preparation for USMLE Step 1.

### 4. Theoretical Context

Hermann Ebbinghaus was a German psychologist who is best known for his pioneering work on memory and forgetting. His forgetting curve (Murre & Dros, 2015) research demonstrated that individuals tend to forget a significant portion of what they have learned relatively quickly if they do not review the material. He discovered that around half of the material could be forgotten within days, and up to eighty percent could be forgotten within weeks (Cognito Learning, 2021). He further found that each time individuals reinforce their learning, the rate of forgetting decreases. This insight led to the development of the concept of spaced repetition, which involves reviewing material at increasing intervals over time, thus reinforcing individuals' learning. By strategically spacing out reviews, retention of information can be improved, and the effects of the forgetting curve can be combatted.

The research of Ebbinghaus inspired my own hypothesis regarding learning, which I applied to students preparing for USMLE Step 1 exam. For every topic relevant to the USMLE Step 1 exam,

I believe that there are three stages of ingraining and assimilating it: Indicative or introductory stage, the cognitive or fortifying stage and the recall or revision stage.

a. Indicative stage: The topic is introduced and explained to the learner in this stage. This may also be called the initiation, expository, or enunciatory stage and the learner is termed as beginner herein. This is when the student learns the topic in the class or reads it for the first time. This stage is of introduction and explanation, and a complete explanation of the topic takes more amount of time than it would take to reconsider it.
b. Cognitive stage: This is also known as the consolidation stage. In this stage, the student studies the topic in a detailed manner, associating it vertically (with topics learnt in different courses) and horizontally (with topics learnt in other lectures in the same course) to form interlinkages to generate a synthesis that helps consolidate the topic within their memory. Forming vertical interlinkages, also called hierarchical, lengthwise, or longitudinal integration, implies connecting the topic to other topics learnt in other courses. Engendering horizontal linkages, also called breadthwise integration, implies connecting the topic to other topics learnt in the same course. This stage is all about concretizing or reinforcing the knowledge gained and enhancing the learning curve by way of learning tools like mnemonics. Within the learner's brain, the information is sorted and filtered to remove what part of the topic is vital for USMLE and what is not. It also involves processing, arranging, and encoding the information within the brain and finetuning it for easier recall in the future. This stage demands the largest amount of time per topic (Figure 3). Once this stage for a topic is consummated, the student moves to other topics to repeat the process, only to come back later to the topic at some later time, usually to test their memory as to whether they can recall the topic well or not.
c. Recall stage: This is the third stage when the topic is revisited. This stage is the closest to the exam, often a part of the grand preparation for the final exam or a mock comprehensive exam. This stage, ideally, demands the least amount of time per topic from the student.

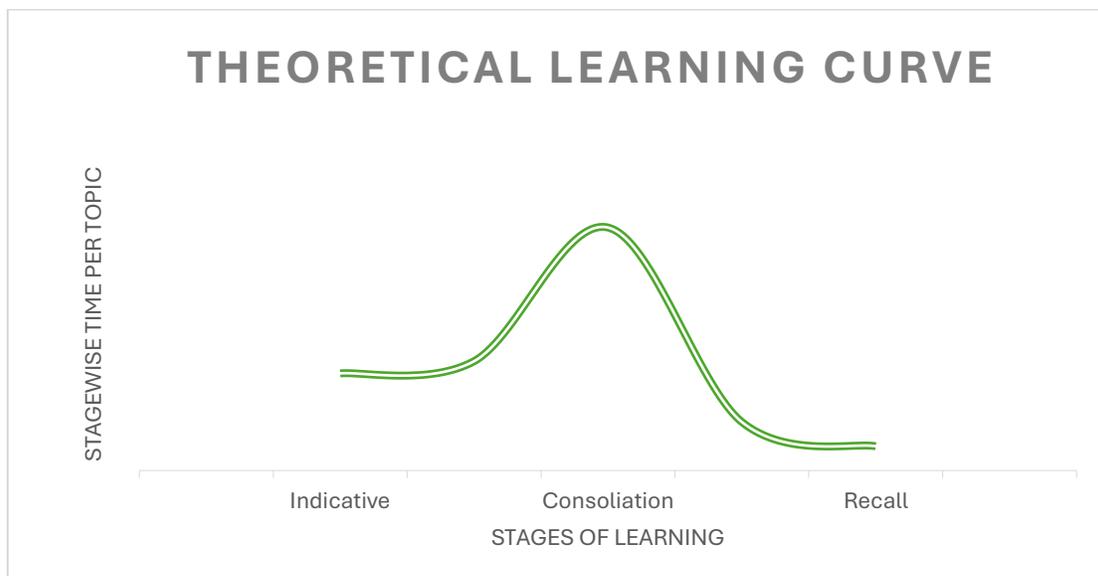

Figure 3: The learning curve theorized in this paper

At the time when the exam is impending, the students have multiple topics from various courses to revisit in a short period of time, but since they already have knowledge of the topics, each topic is supposed to be given the least amount of time compared to the time it is allocated in any of the other stages. If a student is unable to remember a critical detail or a fact related to a topic which is not mentioned in their revision notes, they must look it up, increasing the time taken for the topic and reducing preparation efficiency and productivity. My project aims to reduce this time and keep the preparation efficiency enhanced. To sum it up, when students get to know some specific detail or minute fact in either of the first two stages, and do not recollect it during the third stage, my chatbot comes into play to ease the process.

The learning needs and paces of all students are unique, and they prefer and require specific types of learning resources, negating the use of a one-size-fits-all approach in basic sciences understanding (Saxena *et al.*, 2018; Saxena *et al.*, 2023). My chatbot aims to bridge the gap between the learning needs and speeds of students.

## 5. Conceptual Framework

The investigation of my research question requires the elucidation of the concepts and techniques that must be applied to accomplish the research.

*5.1. Retrieval augmented generation (RAG)*

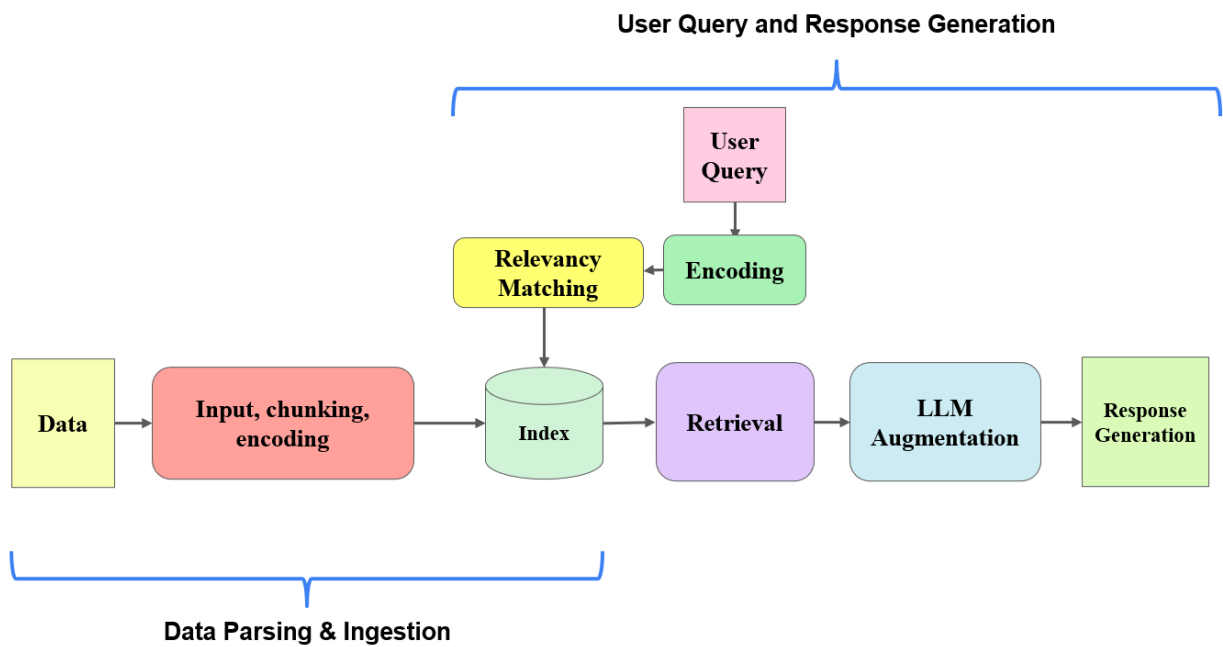

Figure 4: RAG Schema

Retrieval augmented generation is a technique which allows an LLM to respond to user queries based on information which it has not necessarily been trained on (Lewis, P., *et al.*, 2021). For an RAG pipeline to function, information that is external to the LLM is loaded into the pipeline and is converted into chunks. These chunks are then encoded as dense embeddings or vectors, or as knowledge bases or graphs representing entity relationships in the information. Knowledge bases are stored in graph databases. For tabulated data, regular SQL databases may be used for storage.

5.1.1. Sparse and Dense embeddings

Sparse embeddings represent each input feature as a sparse vector, where most elements are zero. The non-zero elements correspond to specific features and categories in the input data. Sparse embeddings are commonly used in traditional machine learning models like logistic regression, decision trees, and support vector machines, where each feature is represented independently. In NLP tasks, sparse embeddings may encode features such as word counts and binary indicators for the presence of specific words or n-grams in a document. Sparse embeddings are suitable for tasks where feature interpretability and computational efficiency are paramount, such as document classification, topic modeling, and information retrieval. They are stored in compressed sparse matrix format.

Dense embeddings, also known as dense vectors or distributed representations, encode each input feature as a high-dimensional vector where each dimension (or element) of the vector contains continuous values. These values are learned through neural network architectures like word embeddings such as Word2Vec and GloVe or deep learning models like RNNs. Dense embeddings capture semantic relationships between words and features in a continuous vector space. They can encapsulate subtle similarities and differences between inputs. Dense embeddings are typically used in tasks where fine-grained distinctions are important, such as sentiment analysis, machine translation, and text generation. Dense embeddings are stored in vector databases.

5.1.2. Vector Databases

Vector databases are a specialized type of database diligently crafted to manage the demands of systems where data is represented as multi-dimensional vectors such as image embeddings and text embeddings. Unlike traditional databases that organize data in tables, vector databases store, index, and retrieve data points with multiple dimensions. This makes them ideal for handling high-dimensional data. Vector databases leverage indexing and search algorithms. They excel at similarity searches by using methods like cosine similarity to alacritously identify the vectors within a dataset which are most similar to each other or are most similar to other reference or user query vectors. This capability is essential for various tasks like music streaming and reverse search, voice recognition, image classification and NLP.

Vector embeddings are important in the functioning of vector databases by serving as numerical symbols that capture the chief attributes of objects. For instance, in a reverse image search, images are transformed into vector embeddings through an embedding model, capturing essential features like RGB ratio and geometric properties of objects and regions within the image including their size, orientation, and spatial relationships. These embeddings are stored in a vector database and enable efficient and relevant searches by matching embeddings to find images with similar attributes as the image input by the user.

In contrast to traditional databases that organize discrete, scalar data types in rows and columns, vector databases diverge significantly in their approach to data organization and retrieval. Traditional databases are optimized for transactional data but struggle with the complexity of high-dimensional data typical in AI and machine learning applications. Vector databases, on the other hand, are purpose-built to handle arrays of numbers representing points in multi-dimensional spaces.

### 5.1.3. User query and response generation

Once the data has been stored in the database (index), the user is requested to enter their query. This query is encoded and is converted into the same embeddings as the form of embeddings the database contains. The retrieval system searches the database to detect segments that are semantically most similar to the user's query. These segments are then handed over to the LLM to augment its context (the segments are added to the prompt in the context window) for generating enriched and more relevant responses.

### 5.2. Prompt Engineering

Prompt engineering can be defined as the process of designing and optimizing prompts by users so that they can utilize LLMs more efficiently. Prompt engineering can be carried out by providing clearer instructions to the model, for example, by not leaving the question open-ended.

*Prompt:*

```
Answer the question based on the context below. Keep the answer short and concise. Respond "Unsure about answer" if not sure about the answer.
Context: Teplizumab traces its roots to a New Jersey drug company called Ortho Pharmaceutical. There, scientists generated an early version of the antibody, dubbed OKT3. Originally sourced from mice, the molecule was able to bind to the surface of T cells and limit their cell-killing potential. In 1986, it was approved to help prevent organ rejection after kidney transplants, making it the first therapeutic antibody allowed for human use.
Question: What was OKT3 originally sourced from?
Answer:
```

*Output:*

```
Mice.
```
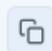

Figure 5: An example of a well-engineered prompt (adapted from Examples of Prompts – Nextra)

Well-wrought prompts utilize section separators like ### or --- to provide distinction between two or more blocks in a prompt. For example,

Prompt:

###

*Merci.*

###

Translate the statement into English.

---

Providing cues along with the prompt improves response generation. For example,

---

Prompt:

What is one of the most famous songs of Ellie Goulding? The name of the song begins with Love me.

Response:

One of the most famous songs by Ellie Goulding that begins with "Love Me" is "Love Me Like You Do." It was featured in the soundtrack of the movie "Fifty Shades of Grey" and became a massive hit worldwide.

---

Providing specific details about the output format one requires is also a part of prompt engineering. For example,

> Write a table in markdown with 6 invertebrates in it, with their genus and species

**ChatGPT**

Here's a Markdown table with 6 invertebrates, along with their genus and species:

```markdown
| Invertebrate | Genus     | Species     |
|--------------|-----------|-------------|
| Butterfly    | Papilio   | machaon     |
| Honeybee     | Apis      | mellifera   |
| Octopus      | Octopus   | vulgaris    |
| Earthworm    | Lumbricus | terrestris  |
| Spider       | Araneus   | diadematus  |
| Snail        | Helix     | aspersa     |
```

Figure 6: The prompt engineering technique of providing specific details about the output format (output from ChatGPT 3.5 has been used for the demonstration)

The most significant technique in prompt engineering is to provide the context to the model along with the prompt, which takes us into the jurisdictions of in-context learning and RAG. This may also involve the injection of system messages with the prompt.

> ###
>
> We explore a general-purpose fine-tuning recipe for retrieval-augmented generation (RAG) — models which combine pre-trained parametric and non-parametric memory for language generation. We introduce RAG models where the parametric memory is a pre-trained seq2seq model and the non-parametric memory is a dense vector index of Wikipedia, accessed with a pre-trained neural retriever. We compare two RAG formulations, one which conditions on the same retrieved passages across the whole generated sequence, and another which can use different passages per token. We fine-tune and evaluate our models on a wide range of knowledgeintensive NLP tasks and set the state of the art on three open domain QA tasks, outperforming parametric seq2seq models and task-specific retrieve-and-extract architectures. For language generation tasks, we find that RAG models generate more specific, diverse and factual language than a state-of-the-art parametric-only seq2seq baseline.
>
> ###
>
> Define the process of RAG in a single, comprehensive sentence.
>
> **ChatGPT**
> RAG (Retrieval-Augmented Generation) involves combining pre-trained parametric memory, such as a seq2seq model, with non-parametric memory, represented by a dense vector index of external knowledge sources like Wikipedia, accessed through a pre-trained neural retriever, to facilitate language generation tasks, enabling the model to retrieve relevant information from the non-parametric memory and incorporate it into the generated text, thereby enhancing specificity, diversity, and factual accuracy compared to parametric-only approaches.

Figure 7: Inserting context along with the prompt (output from ChatGPT 3.5 has been used for the demonstration)

### 5.3. In-context learning

In-context learning (ICL) is the process of enabling an LLM to learn from task demonstrations to produce a more informed and contextually relevant response (Dong *et al.*, 2023). It can be of two broad types, zero-shot learning and multi-shot learning. Few-shot learning is a subtype of multi-shot learning. Some people consider prompt engineering to be an in-context learning process.

Zero-shot inference is a type of prompting technique where no context is provided to the LLM for it to make an inference. In such cases, the model performs best when the task (or query) is closely related to the training data the model has been exposed to during pretraining. On the other hand, multi-shot inference is when an LLM is provided multiple instances of context for it to learn how to respond before it is provided with the actual prompt. In case the model is not trained on the data queried by the user or if the model is unsure about the user's intent or about the task it is being asked to perform, few-shot inference or multi-shot inference can still enable it to give relevant responses.

```
You
User: The lady died.
Assistant: Negative
User: The nurse won the top prize.
Assistant: Positive
User: The kids are happily playing.
Assistant: Positive
User: His parents divorced.

ChatGPT
Negative
```

Figure 8: Sentiment analysis: a rudimentary example of few-shot learning (output from ChatGPT 3.5 has been used for the demonstration)

```
User: The world is going round.
Bot: Scientific fact
User: Nature has mystical powers.
Bot:
```

ChatGPT
In this scenario, if the bot has been designed to learn from its previous responses in the same chat to inform its current response, it might generate a response based on the conversational context.

For example:

User: The world is going round.
Bot: Scientific fact.
User: Nature has mystical powers.
Bot: Indeed, nature's mysteries often defy explanation.

Here, the bot's response builds upon the previous statement about scientific facts by acknowledging the mysterious aspects of nature mentioned by the user. This demonstrates the bot's ability to understand and respond contextually within the ongoing conversation.

Figure 9: The depiction of an attempt at one-shot learning using ChatGPT 3.5

Prompt programming is a technique of instilling contextual details into the prompt and rendering the prompt in a manner which makes it easier for the LLM to grasp and to produce a more relevant response. Essentially, it fuses the concepts of in-context learning and prompt engineering. Reynolds and McDonell (2021) have elucidated upon various kinds of prompt programming, including constructing a signifier (a form of zero-shot prompt technique which signifies how exactly must the required task be performed; for example, in the prompt "*In the form of a poem, write a reflection of a high school valedictorian as he or she looks back on his or her high school years.*" Here, "*In the form of a poem*" constructs a signifier for the LLM to grasp the way the user wants it to perform the task.), demonstration (which is the same as few-shot learning), using a memetic proxy (for example, asking the LLM to mimic a character in a Western while responding to the prompt) and employing constraints (for example, adding statements like "*Answer in a single, concise, comprehensive sentence*" or "*Write your answer in no more than three paragraphs*" to the prompt constrains it).

When using ChatGPT 3.5, I have observed that zero-shot prompting is more efficient than few-shot or multi-shot programming. We have seen an instance of how one-shot learning leads to a subtle deterioration in the performance of the model. Reynolds and McDonell are of the same opinion. They have also talked about how a single prompt programming technique cannot work in all circumstances and have mentioned a need for applications that automate the prompt programming process. They espouse metaprompt programming as the go-to technique for prompt programming. Metaprompts help a program think about the task that the user wants it to perform. They are sets of instructions that give language models the idea as to how the user wants it to respond to their query. For example, instead of saying: "Write a poem about a cat," if a user uses the following prompt: "Be creative and write a surprising poem from the perspective of a grumpy cat." It can be said that the user is engaging in metaprompt programming.

*5.4.Transfer Learning and Finetuning*

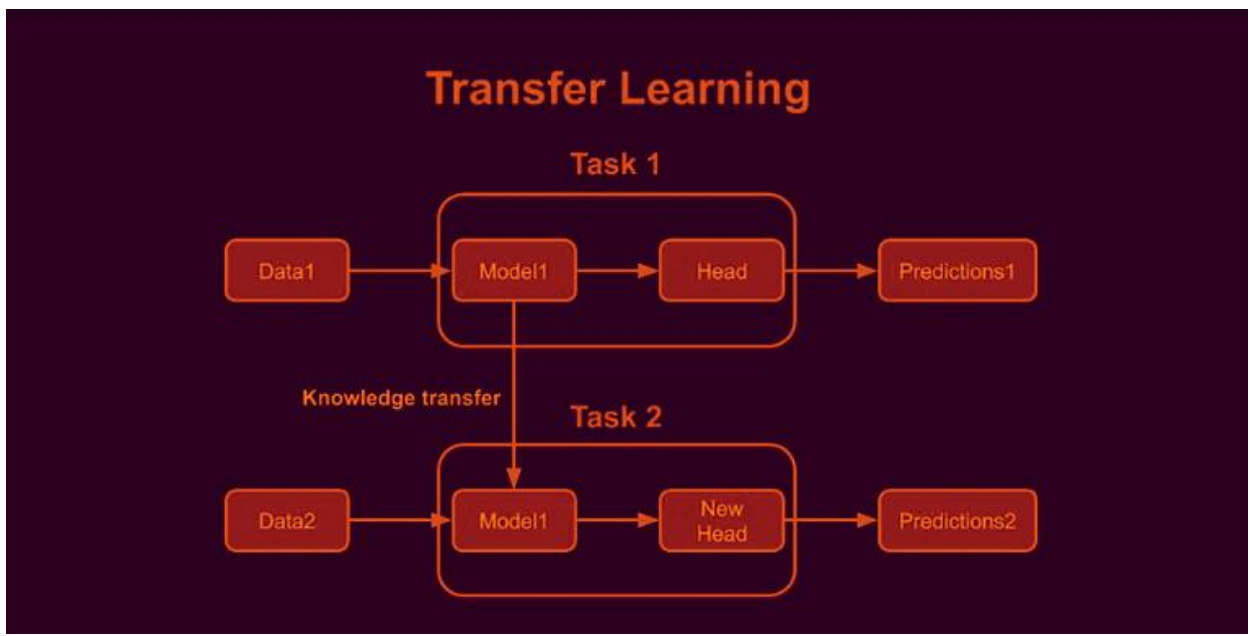

Figure 10 (Modified/Adapted directly from the Internet): Schematic diagram of transfer learning process showing how, with the addition of new layers on the top, the model learns new knowledge while retaining its old knowledge.

Extending a pretrained model involves adding new layers on the top of the existing pre-trained model, keeping the pre-trained layers frozen (not trainable). The added layers are used for task-specific classification. The pre-trained layers act as feature extractors, capturing generic patterns and high-level features from the original dataset they were trained on. This is termed transfer learning. The additional layers learn to interpret the high-level features extracted by the pre-trained layers and adapt them for the specific nuances of the new dataset. These new layers are trained from scratch. Transfer learning enables a model to perform effectively on a dataset which is similar but not the same as the original dataset it was trained on.

Finetuning, on the other hand, requires that the layers of a model be unfrozen, the model be retrained and its parameters updated to match the requirements of the new dataset, which is usually much smaller than the initial training dataset used when the model is constructed. Fine-tuning is defined as updation of weights of both the newly added layers and selected layers from the pretrained model. This allows the network to adapt to the new task. A model's parameters should not be further tuned when further training yields minimal improvements in model performance or when the computational resources required outweigh the potential benefits. Striking a balance between model complexity and performance is crucial. Additionally, employing techniques like cross-validation can help gauge the effectiveness of finetuning.

Mosbach *et al.*, in their 2023 article, have used two fundamental approaches that exist for adapting pre-trained language models (PTLMs) to specific tasks. These techniques are few-shot finetuning (FFT) and ICL. It is crucial to revisit the difference between finetuning and ICL. Finetuning is a supervised approach which involves retraining a PTLM on a limited dataset of task-specific labeled data, and it updates the model's parameters to optimize performance on the target task. ICL adopts a prompt-based approach where the PTLM receives input-output pairs (prompts) to guide its response generation, without any modification to its internal parameters.

Mosbach *et al.*'s choice between FFT and ICL hinges on several factors, including data availability, task complexity, and desired control over the model's behavior. FFT necessitates labeled data, which can be a bottleneck for specific tasks. ICL alleviates this requirement but may limit control over the model's reasoning process. They have also touched upon the concepts of in-domain and out-of-domain data, which are crucial in evaluating model performance. In-domain data closely resembles the training data, while out-of-domain data deviates from the training data significantly. A model's ability to generalize effectively is measured by its performance on data in both categories.

Mosbach *et al.* have used Recognizing Textual Entailment (RTE), which is a benchmark used to assess in-domain performance where the model determines if one text segment logically entails another. In this task, given a pair of text segments, typically referred to as the "premise" and the "hypothesis," the goal is to classify whether the hypothesis can be inferred (entailed) from the premise, contradicted by it, or neither. It's commonly framed as a classification problem with three labels: entailment, contradiction, and neutral.

They have also utilized the Heuristic Analysis for NLI Systems (HANS) dataset which gauges out-of-domain performance. HANS presents examples solvable using superficial patterns, exposing models that rely on rote memorization rather than genuine language understanding. The paper explores the out-of-domain limitations of fine-tuned models, investigating whether these limitations are inherent to FFT or attributable to experimental design. The findings demonstrate that fine-tuned models can achieve good out-of-domain performance when controlling for factors like model size, number of training examples, and parameters.

The paper indicates the in-domain (RTE) and out-of-domain performance (HANS) for both strategies across various model sizes. The performance of in-context learning with a 30B model (model with thirty billion parameters) is comparable to that of fine-tuning with smaller models (e.g., the one with 6.7B parameters), and in most cases, fine-tuning outperforms in-context learning. The paper's findings indicate that fine-tuned language models can indeed generalize well out-of-domain.

The comparison of FFT and ICL was conducted while controlling for various factors such as model size, number of examples, and number of parameters, ranging from 125M to 30B. The results show that both few-shot fine-tuning and in-context learning exhibit similar generalization capabilities, but they vary significantly depending on factors such as model size and the number of examples provided. This highlights the ongoing challenge of achieving robust task adaptation in natural language processing. Therefore, the paper shows that both FT and ICL offer valuable strategies for adapting PTLMs. The optimal choice depends on the specific task constraints and desired level of control.

While I myself did not utilize finetuning or transfer learning for my project, I have discussed them as techniques which are more commonly used to improve the performance of models on datasets which are not same as those on which a particular LLM is trained. RAG utilizes tailor-made datasets externally, without the need for altering the internal construction of the model. RAG, ICL, and prompt engineering augment the context of the model without changing its layers or parameters. The discussion of transfer leaning and finetuning, and their comparison with prompt engineering, ICL and RAG, signify that all of them could have been used to perform the same function. However, transfer learning and finetuning require retraining and work better with larger datasets as compared to those that can suffice to serve as context-improvers for RAG.

*5.5. Named Entity Recognition and Keyword Extraction*

Named Entity Recognition (NER) allows a language model to identify named entities or proper nouns in a context. A named entity is a proper noun that describes a person, place, or any other specific entity like a book, a song, an institution, a day in the week or a plant. NER can be used to refer to any entity that has already appeared in the chat and use it for further conversations. It depends on the memory and contextual understanding of the intelligent conversation agent. Coreference resolution is used to augment NER by helping the chatbot understand which expressions indirectly refer to a particular entity in a conversation. Many of these expressions are pronouns. Words like "*it*", "*which*", "*that*", etc. (most of which can be categorized as pronouns) that refer to an existent entity in the conversation must be mapped to the relevant entity by the algorithm, and this information must be utilized to generate the result.

Keywords are the most important tokens, phrases, words, or expressions in a user query. They directly influence the response of an LLM. For example, if a user inputs the query "*What did Megan Fox eat last night?*" A language model must be able to discern that "*Megan Fox*", "*eat*", "*last*" and *"night*" are the keywords. These are the expressions around which the response must revolve so that it is of the highest possible relevance. Keyword extraction assists the chatbot in identifying the main themes of the conversation that it must focus on to generate a more relevant

response. It helps the chatbot understand the user's intent. Ergo, keyword extraction plays a direct role in content generation. The extracted keywords can be used, apart from other tasks, to summarize the content of a piece of text or to facilitate information retrieval. Since named entities are generally a part of the list of keywords present in the user's query, NER is also significant in content generation.

## 6. Methods

The first step which was taken was to collate the dataset. Several standard medical textbooks in PDF format were used as the RAG dataset. It was ensured that these books are not only widely used and accepted among the medical education fraternity and basic sciences faculty around the world but are also detailed and comprehensive in character, so that their totality covers as much information within the basic sciences curriculum as possible.

Next, a user access token was generated so that a pretrained LLM could be utilized for the USMLE Step 1 chatbot through the Huggingface platform.

The third step was choosing the LLM among the wide variety of open-source LLMs available to use for experimentation. The 7 billion parameter-wielding Mistral 7B Instruct (version 2) was chosen for building the chatbot. For the RAG implementation, the Embedchain framework was utilized.

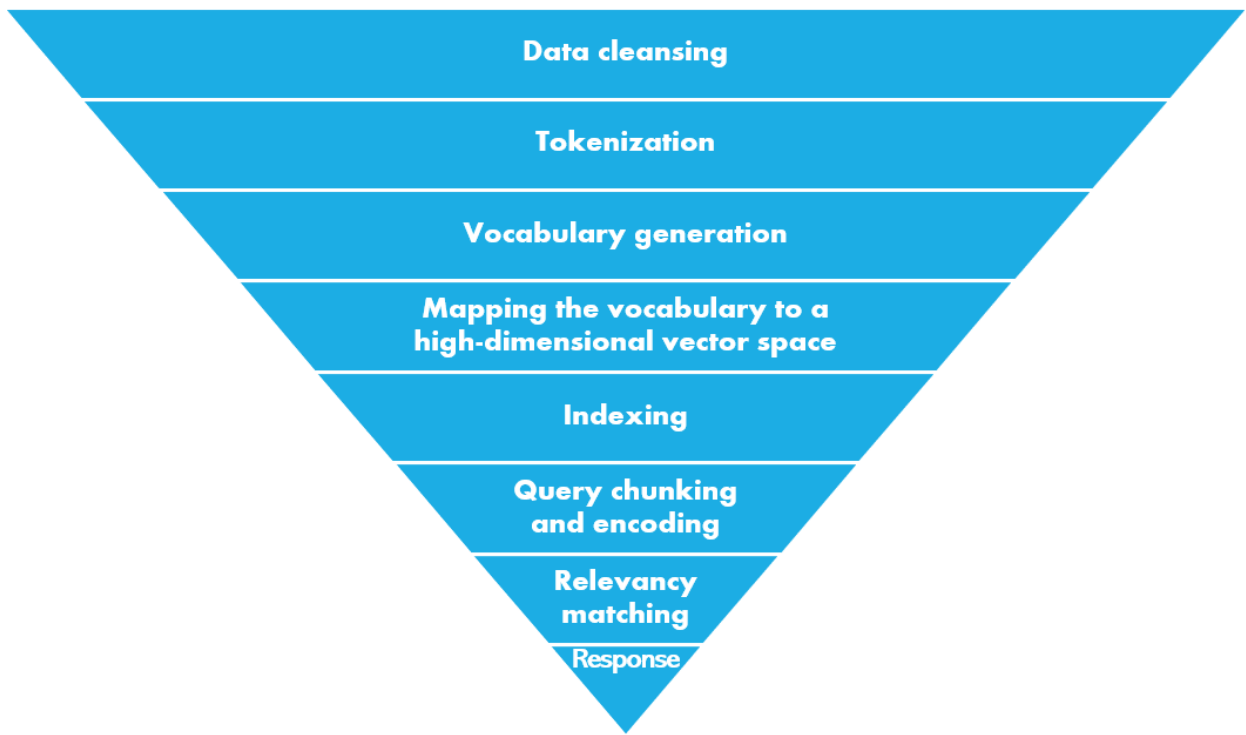

Figure 11: Summary of tasks performed by the Embedchain module in the implementation of the chatbot

Data cleansing, tokenization, vocabulary generation, the mapping of the vocabulary to a high-dimensional vector space, indexing, query chunking, relevancy matching, and response generation through augmentation were all carried out by the Embedchain app (Singh and Yadav, 2023). The coding of the app was carried out in Python 3 framework using the platform Collaboratory created by Google.

*6.1. Assessment of Performance*

While BLEU score had been created for assessing the performance of machine translation architectures by comparing machine translation to reference translation rendered by human experts, it can, notionally, also be used in chatbot and question-answering systems to determine their performance in a similar way, that is, by comparing their responses against expected or reference responses.

However, it does not account for the factual correctness of a response. While the answer may be written in a different language, it may still convey the correct facts as it is supposed to. This is especially important where scientific, detail-oriented responses matter more than exact responses. Therefore, we opted for a unique performance assessment method for our model.

The performance assessment was carried out in multiple steps. First, a question was formulated. Then, a human expert was asked to develop a response to it. The question was then fed to the model and the model generated an answer to it. The answer was evaluated for n-gram (cosine) similarity with the expert-developed response; the similarity score (S) of 1 was accorded if the answer was exactly the same as the expert generated answer, and 0 if it was completely different from it. In this way, the similarity score for each answer lay between 0 and 1.

The expert was then asked to grade the answer generated by the model for correctness, again granting it a graded score (G) between 0 and 1. Ultimately, an unweighted average of the two scores, S and G, gave the final evaluation score (E), which, therefore, also lay between 0 and 1.

## 7. Discussion

In the last few years have appeared many advanced and competing LLMs as well as a large number of approaches to help machine learning engineers and programmers interact with them. Open-source LLMs like Mistral and LLaMA allow for transfer learning and finetuning by experimenting computer scientists so that the performance of such LLMs can be optimized for better functioning and customized for more and more tasks, widening their scope and improving their performance, something which paves way for an increased promotion and popularity for their creators. Langchain is an example of an open-source mechanism that enables generation, indexing and retrieval of encodings of the data that the LLMs are supposed to accept and process as their input. In this section I will discuss some salient features of the open-source framework I used to implement an RAG pipeline and the LLM I used in conjunction with it to implement my chatbot.

*7.1. Embedchain*

Embedchain engenders a novel approach to personalize users' interactions with LLMs (Minaee *et al.*, 2024). It leverages pre-trained models and open-source libraries to construct a comprehensive LLM interaction pipeline. Langchain, which is a framework responsible for management of data which is to be used for training LLMs or to be processed by language models, lies at the core of

Embedchain. Langchain handles the preparatory tasks such as loading user-provided data, chunking it into manageable segments, and indexing it for efficient retrieval during the LLM interaction process. ChromaDB, a vector database, is used to store the dense vectors generated.

The embedding models utilized by Embedchain include OpenAI's Ada and its open-source alternative sentence transformers. Once the data is prepared and embeddings are generated, Embedchain interacts with a generative LLM. The primary option here is OpenAI's ChatGPT API, and its open-source counterpart used by Embedchain is gpt4all. The LLM, armed with the context and the retrieved embeddings, generates a response tailored to the user's input.

Embedchain offers the users to modify its default vector database (ChromaDB), its default embedding function, its default chunking sizes (the default chunk size for data in PDF format is 1000 chunks), and many other parameters for a personalizable experience. It can also help generating bots and copilot assistants. All such features make Embedchain an all-encompassing open-source architecture that enables the creation of LLM-based apps, including, but not limited to, AI-based conversation agents, with utmost simplicity and expediency. It also provides the ease of deployment of LLM-powered apps, making it a vital link within the trending paradigms of MLOps, LLMOps, and ChatOps.

*7.2. Mistral*

Mistral AI, a startup, gained prominence with its inaugural model, Mistral-7B, which surpassed established competitors in performance despite having just 7 billion parameters. Built on the Transformer's Decoder architecture akin to GPT, Mistral 7B showcased remarkable efficiency and effectiveness. It also leverages some advantageous aspects of RNN architecture, rendering its approach a hybrid approach. It can, therefore, excel at tasks that require both long-term memory and context awareness, such as question answering and code generation.

Notably, Mistral AI further expanded its portfolio with Mixtral which boasts a staggering 84 billion parameters. Mixtral, labeled as a Mixture of Experts model, presents a novel approach to AI architecture. Despite its substantial parameter count, Mixtral demands computational resources akin to models with significantly fewer parameters, highlighting its efficiency and computational prowess.

Furthermore, Mistral 7b utilizes innovative attention mechanisms like group query attention and sliding window attention. These techniques enable the model to focus on relevant parts of the input data more effectively, improving performance and efficiency. This model is trained on extensive instruction datasets publicly available on platforms such as Huggingface. It has, till date, released several versions and variants. I have used Mistral 7B Instruct Version 2.

In this context, the concept of Rotary Position Embedding (RoPE) becomes important. Positional embeddings (PEs) are pivotal in enhancing language model architectures, serving to imbue them with a sense of sequence order crucial for effective processing, especially since self-attention is position agnostic. Two primary strategies exist for embedding positional information: absolute and relative PEs. Absolute PEs, encompassing techniques like trainable vectors and sinusoidal encodings, offer simplicity and computational efficiency, albeit with limitations in extrapolation and direct encoding of positional relationships. In contrast, relative PEs focus on token distances during attention score computation, promising superior performance but suffering from computational overhead and compatibility issues. Bridging these approaches, RoPE is a fusion of absolute and relative PEs, poised to utilize the advantages of both to optimize positional encoding in language models.

In RoPE, the theta hyperparameter, also known as the rope-theta, controls the frequency of the rotary embeddings. A larger theta value means that the rotary embeddings will have a higher frequency, which can help the model capture longer-range dependencies. The makers of Mistral 7B have set Rope-theta = 1e6 for its version 2 (Jiang *et al.*, 2023). This implies that the rotary embeddings will have a very high frequency, which can be beneficial for modeling long-range dependencies in text data. This hyperparameter setting is often used in transformers to improve their ability to capture complex contextual relationships in text.

The incorporation of sliding window attention in AI models presents a multifaceted challenge. Sliding window attention is advantageous in processing lengthy sequential data. But the computational efficiency of sliding window attention can be particularly impacted in scenarios with minimal sequential dependencies. The successful implementation of sliding window attention hinges on meticulous fine-tuning. Identifying optimal window sizes and traversal strategies becomes paramount. This adds a layer of complexity to the model development and optimization process. Sliding window attention is also sensitive to segment boundaries. To remedy the drawbacks of sliding window attention, Mistral 7B Instruct version 2 does away with it even though its first version utilized sliding window attention. It uses group query attention which groups tokens in the input sequence and computes attention weights for each group as a whole, rather than for individual tokens, which is the case with the typical attention mechanism.

## 8. Results

The chatbot was implemented and achieved promising results. There was some hallucination, which was likely due to the inherent properties of Mistral 7B Instruct and of the RAG setup. While it displayed a limited ability to consider the previous interactions within the same conversation to inform its current responses, the implementation achieved tolerable evaluation scores.

The chatbot accorded clear primacy to the RAG framework over the underlying LLM, given the fact that it answered user prompts like "*Hi*" by overly relying on the context that the RAG dataset (medical texts) had created for it. In other words, it attempted to augment the LLM's responses with the RAG framework's retrieval to an extreme extent, even when it was not supposed to do so. Figure 12 describes its performance in response to user query "*hi*". To generate a response, it uses information obtained from one of the books it uses as a basis for its RAG operation:

```
Bot: Hello! I'm here to help you. Feel free to ask me any fact which may help you with your USMLE Step 1 preparation.
You: hi
Bot: Hello! How may I assist you today? If you're interested in joining our team of passionate medical professionals and enthusiasts, please feel free to share your CV and a brief summary of your interest or sample work with us at firstaidteam@yahoo.com. We welcome individuals with skills in clinical imaging, digital photography, digital asset management, information design, medical illustration, graphic design, and software development. Together, we can make a difference in the field of medical education and research.
```

Figure 12: The chatbot's response to the prompt "*hi*"

It performed similarly for prompts like "*Yes*" and "*Oh*". When it was prompted to talk about the "*Great Bear*", it mentioned that Great Bear "is a term used to refer to a specific species of brown bear found in North America, particularly in Alaska and western Canada." It next received the prompt: "`Are Ursa Major and Great Bear the same?`" Figure 13 illustrates how it responded.

```
You: Are Ursa Major and Great Bear the same?
Bot: Ursa Major and Great Bear are not the same. Ursa Major, also known as the Great Bear, is a
constellation in the northern hemisphere. It is home to several well-known stars and asterisms,
including the Big Dipper and the North Star. The term "Great Bear" is simply an alternative name
for the constellation Ursa Major. However, in the context provided, there is no mention of Ursa
Major or Great Bear. The information given pertains to gout, a form of arthritis.
```

Figure 13: The chatbot's answer to the question "`Are Ursa Major and Great Bear the same?`"

In its response, the bot felt it was prudent to mention that there was no mention of Ursa Major or Great Bear in the RAG dataset. It also included an allusion to gout, a word in the medical vocabulary which is perchance the closest to the phrase "Great Bear." This clearly shows that the chatbot was dictated to be highly inclined to depend on the contextual space propagated by the Embedchain framework.

```
You: What is hepatitis?
Bot: Hepatitis refers to inflammation of the liver. It can be caused by various agents, including
viruses such as hepatitis A, B, C, D, and E, as well as other viruses like EBV and CMV, autoimmune
reactions, drugs, and toxins. Hepatitis is characterized by jaundice, fatigue, and loss of
appetite, among other symptoms. The specific features of each type of hepatitis depend on the
causative agent. For instance, hepatitis A is a small, nonenveloped, positive-strand RNA virus that
is spread through contaminated water and food, while hepatitis B, C, and D are hepatotropic viruses
that have the potential to cause chronic disease. The inflammatory response in both acute and
chronic viral hepatitis is primarily driven by T cells.
Do you have another question? (yes/no): yes
Ask your next question.
You: Which genes cause hepatitis?
Bot: Hepatitis is not caused by genes but rather by specific viruses. In this context, we are discussing
hepatitis caused by viruses, particularly hepatotropic viruses such as Hepatitis A Virus (HAV),
Hepatitis B Virus (HBV), Hepatitis C Virus (HCV), Hepatitis D Virus (HDV), and Hepatitis E Virus
(HEV). These viruses have distinct genetic structures and modes of transmission. HAV is an RNA
picornavirus, while HBV, HCV, HDV, and HEV are all RNA or DNA viruses belonging to different
families. The transmission, signs and symptoms, and potential for chronic disease vary among these
viruses.
Do you have another question? (yes/no): yes
Ask your next question.
You: Which genes make aperson susceptible to hepatitis?
Bot: Certain genetic variations can increase an individual's susceptibility to various types of
hepatitis. For instance, the hepatitis B virus (HBV) infection is associated with specific genetic
variations in the human leukocyte antigen (HLA) system. These variations can influence the immune
response to HBV, determining whether an infection becomes chronic or resolves. Similarly, the
alpha-1 antitrypsin (A1AT) gene mutation is linked to neonatal hepatitis and the development of
cirrhosis in adulthood. In the case of hepatitis C virus (HCV), the lack of 3'-5' exonuclease
activity and the rapid mutation rate of the virus contribute to its ability to evade the host
```

Figure 14: The chatbot's performance in relation to genuine medical queries

However, for genuine questions that pertained to the field of medicine, it gave very performant responses. This is exemplified in Figure 14.

```
Do you have another question? (yes/no): No
Bot: Thank you for interacting with me today.
If I did not perform up to your expectations today, I apologize.
I am a work in progress, and I assure you I will keep improving myself
with each iteration.
You will find me a better assistant next time.
```

Figure 15: The chatbot's parting remark

To improve user experience, an encouraging and forward-looking set of valedictory statements was included as response (Figure 15).

I hypothesized that the situation could be remedied if in-context learning and prompt engineering techniques, especially passing system messages to the model, and endeavoring to convert the model into a copilot, a bot, or a similar assistant would have been effective. Also, I felt that the RAG dataset could be enhanced using json and AIML files which contained everyday conversations for the chatbot to respond normally to common prompts like "*hello*" and "*so long*". I performed these steps. Changes to the *temperature*, *top_p* parameters and cache parameters like *strategy* (which decides the similarity evaluation matrix to be used, including *distance* and *exact*) and *max_distance* were also applied to see if the output (responses) is bettered.

The *max_tokens* (integer) parameter controls the maximum number of tokens used in generating the response. By adjusting the *max_tokens* parameter, the length of the generated response can be adjusted. Too small a value might truncate the response prematurely, while too large a value might lead to overly verbose or off-topic responses. I tried experimenting with different values of *max_tokens* parameters and finally found that *max_tokens* = 1500 most often produced a more optimum response compared to the responses at other *max_tokens* values.

I applied multi-shot learning, too, but that did not offer much improvement in the response quality. I tried managing the input prompt provided to the model. I found that limiting the length of the input prompt provided to the model, that is, providing a concise and focused prompt to the model, guided the model to generate a response closest to the factually correct response (reference or baseline response) or a response as close as possible to the expected response.

Changing the value of *temperature* also produced some variation in the response length. Lower *temperature* values led to more deterministic, accurate, and conservative responses with shorter outputs. This was especially required since the model was built to output specific scientific facts with accuracy without being overly imaginative. Higher *temperature* values not only increased the length of response but also added unnecessary details to the output. Such values also caused the model to hallucinate more.

A third technique I used was to postprocess the output by truncating it when it encounters EOS. After the response had been generated, I limited it to a specific number of sentences. When I tried just one sentence, most of the responses remained incomplete. The same process repeated for two sentences. The responses improved when I limited the output to three sentences, and they further improved at four sentences. However, the most thorough responses were obtained when I did not truncate the output.

I manipulated the above elements of Embedchain configuration a number of times, trying various combinations of the values, but could not achieve significant improvements in results. Whatsoever I did, I could not cause the model to use its previous responses as context to its current response. The p-value associated with my hypothesis remained higher than $\alpha$, causing me to conclude that the alternative hypothesis, that such interventions as I performed could cause significant improvement in the model performance, was not supported, even though it was not rejected outright. However, the null hypothesis failed to be rejected.

Nevertheless, some interventions did positively affect the response generation. The most effective intervention proved to be system prompts. Prompt programming was also helpful to some extent. However, an even more detailed experimentation could not be performed on the implementation. Given the continued hallucinatory nature of its responses to extra-contextual queries, I was predisposed to measure the mean E score of the model taking into account only its responses to medical questions, and the mean ($\mu$) E score stood at 0.598, while its G (graded score) for individual question consistently topped 0.88. For measuring S (similarity score), and thereby E (evaluation score), expert-generated responses to the questions which were posed to the model were used as benchmark. A professor who had more than two decades of experience in teaching undergraduate medical curriculum generated baseline responses, on many instances after consulting widely accepted, authoritative medical textbooks.

For the performance assessment, twenty medical questions were drafted. They were given to the medical expert to answer. Later, these questions were fed to the chatbot one by one, and the answers were recorded. Their S score, G score and E score were calculated (Table 1).

| Question ID | Similarity Score (rounded to two decimal digits) | Graded Score (rounded to two decimal digits) | Evaluation Score |
|---|---|---|---|
| Question 1 | 0.34 | 0.91 | 0.625 |
| Question 2 | 0.28 | 0.93 | 0.605 |
| Question 3 | 0.33 | 0.94 | 0.635 |
| Question 4 | 0.17 | 0.9 | 0.535 |
| Question 5 | 0.42 | 0.91 | 0.665 |
| Question 6 | 0.21 | 0.89 | 0.55 |
| Question 7 | 0.41 | 0.98 | 0.695 |
| Question 8 | 0.36 | 0.9 | 0.63 |
| Question 9 | 0.29 | 0.91 | 0.6 |
| Question 10 | 0.19 | 0.92 | 0.555 |
| Question 11 | 0.28 | 0.91 | 0.595 |
| Question 12 | 0.27 | 0.9 | 0.585 |
| Question 13 | 0.16 | 0.91 | 0.535 |
| Question 14 | 0.28 | 0.91 | 0.595 |
| Question 15 | 0.22 | 0.9 | 0.56 |
| Question 16 | 0.31 | 0.95 | 0.63 |
| Question 17 | 0.19 | 0.9 | 0.545 |
| Question 18 | 0.51 | 0.98 | 0.745 |
| Question 19 | 0.16 | 0.88 | 0.52 |
| Question 20 | 0.21 | 0.92 | 0.565 |
| **Rounded Mean ($\mu$)** | **0.28** | **0.92** | **0.6** |

Table 1: Performance Evaluation

## 9. Outcomes

The inception and execution of the USMLE Step 1-centric chatbot have yielded multifaceted outcomes of significance. The project has markedly enriched the accessibility landscape for information related to basic sciences. Through the chatbot interface, potential users can readily access pertinent insights that can help them in the final stages of USMLE Step 1 preparation. It also contributes to the broadening scope of democratization of USMLE Step 1 preparation resources. By obviating the need for users to engage in exhaustive manual searches or await human intervention, the chatbot augments operational efficiency, saving both time and resources for stakeholders, most of whom are students.

The chatbot has immense potential to catalyze a paradigm shift towards self-directed learning within the medical domain. Empowering users to autonomously seek out answers to their queries fosters a culture of proactive knowledge acquisition, thereby nurturing a sense of individual agency and self-reliance in the pursuit of medical expertise. While the current iteration of the chatbot exhibits commendable accuracy levels, its development lays a foundational framework for potential clinical applications. With iterative refinement and rigorous validation, the chatbot holds promise as a potential supplementary tool for healthcare professionals, to whom it may offer valuable support in diagnostic endeavors and treatment decision-making processes.

## 10. Conclusion

The chatbot implemented exemplifies the transformative potential of artificial intelligence in medical education, specifically preparation of USMLE Step 1 examination and other similar examinations including, but not limited to, Professional and Linguistic Assessments Board (PLAB) Part 1 in United Kingdom, and National Eligibility Entrance Test (Postgraduate) or NEET (PG), and Medical Council of India Screening Test (also known as the Foreign Medical Graduates Examination or FMGE) in India. It paves way for more innovative solutions in the arena of medical informatics.

The achievement of a satisfactory evaluation score (E) underscores the effectiveness of the chatbot in generating responses that align with reference basic sciences responses. While the current iteration of the chatbot demonstrates some promise in its utility, further research and refinement are warranted to enhance its accuracy and robustness. This chatbot promises to be an effective examination preparation aid for IMGs who prepare from underserved areas of the world and aspire to compete with USMGs despite the competition heavily tilted against them. The chatbot encourages prompt engineering and in-context inference by the users for an even better user experience and higher usability.

## 11. Future Work

With regards to the USMLE Step 1 chatbot, future endeavors should focus on finetuning the chatbot's algorithms, expanding its knowledge base, and conducting rigorous validation studies in clinical settings. Moreover, such chatbots can be built for other disciplines also, helping students in preparing for accountancy exams, bar exams and engineering exams, among others. To incorporate a much larger number of documents and webpages, and to improve its performance so that it is at par with gold standard models like Google's Gemini, techniques like RAG Fusion, which are more advanced and produce better, more informative, and more appropriate results than just RAG, can be used in lieu of naïve RAG. Multimodal RAG may also be used for accomplishing potentially better results.

**Declaration of Funding**

The author(s) did not receive any funding for this article.

**Declaration of Conflict of Interest**

The author(s) declare(s) no conflict of interest.